\documentclass[superscriptaddress,
twocolumn,
floatfix,aps,prb,amsmath,amssymb,
]{revtex4-1}

\usepackage{graphicx}
\usepackage{dcolumn}
\newcolumntype{d}[1]{D{.}{.}{#1}}

\usepackage[colorlinks=true]{hyperref}
\usepackage{multirow}
\usepackage{bm}
\usepackage{times}
\usepackage{color}
\usepackage{mathcomp}
\usepackage{siunitx}
\usepackage{eqnarray}
\usepackage{placeins}

\newcommand{\SrIrO}{$\rm Sr_2IrO_4$}

\newcommand{\Tc}{\ensuremath{T_\text{c}}}
\newcommand{\ttg}{\ensuremath{t_\text{2g}}}

\newcommand{\jeff}{\ensuremath{J_{\text{eff}}}}
\newcommand{\Ueff}{\ensuremath{U_{\text{eff}}}}
\newcommand{\kv}{\ensuremath{\mathbf{k}}}

\begin{document}

\title{Optical anisotropy of the \texorpdfstring{\jeff{} = 1/2}{jeff=1/2} Mott insulator \texorpdfstring{Sr$_2$IrO$_4$}{Sr2IrO4}}

\author{D.~Pr\"opper}
\author{A.~N.~Yaresko}
\author{M.~H\"oppner}
\author{Y.~Matiks} 
\affiliation{Max-Planck-Institut f\"ur Festk\"orperforschung, Heisenbergstra{\ss}e~1, D-70569 Stuttgart, Germany}
\author{Y.-L.~Mathis}
\affiliation{Synchrotron Facility ANKA, Karlsruhe Institute of Technology, 76344 Eggenstein - Leopoldshafen, Germany}
\author{T.~Takayama}
\author{A.~Matsumoto}
\author{H.~Takagi}
\affiliation{Max-Planck-Institut f\"ur Festk\"orperforschung, Heisenbergstra{\ss}e~1, D-70569 Stuttgart, Germany}
\affiliation{Department of Physics, University of Tokyo, Hongo, Tokyo 113-0033, Japan}
\author{B.~Keimer}
\author{A.~V.~Boris}
\affiliation{Max-Planck-Institut f\"ur Festk\"orperforschung, Heisenbergstra{\ss}e~1, D-70569 Stuttgart, Germany}

\date{\today}

\begin{abstract}
We report the complex dielectric function along and perpendicular to the $\rm IrO_2$ planes in the layered perovskite 
 \SrIrO{} determined by spectroscopic ellipsometry in the spectral range from 12\,meV to 6\,eV. Thin high quality single crystals were stacked to measure the $c$-axis optical conductivity. In the phonon response we identified 10 infrared-active modes polarized within the basal plane and only four modes polarized along the $c$-axis, in full agreement 
 with first-principle lattice dynamics calculations. We also observed a strong optical anisotropy in the near-infrared spectra arising from direct transitions between $\rm Ir\ 5d$ \ttg{} \jeff=1/2 and \jeff=3/2 bands, which transition probability is highly suppressed for light polarized along the $c$-axis. The spectra are analyzed and discussed in terms of
relativistic LSDA+U band structure calculations. 
\end{abstract}

\pacs{78.20.Ci, 63.20.-e, 71.70.Ej, 71.20.-b}
\maketitle

\section{introduction}

A rich variety of electronic ground states of transition metal oxides (TMOs) emerges from strong electron correlations and cooperative phenomena with competing interactions, including the on-site Coulomb repulsion $U$,  crystal-electric field (CEF), and spin-orbit coupling (SOC). The transition from elements with 3d via 4d to 5d valence orbitals progressively results in larger single particle band width $W$, reduced $U$, and enhanced SOC. TMOs of the type $\rm(La,Sr)_2MO_4$, where $\rm M = Cu$ (3d), $\rm Ru$ (4d), or $\rm Ir$ (5d), allow one to consider the magnitudes of these interactions as variable parameters which can significantly influence the electronic structure within the same layered perovskite '214' structure. $\rm La_2CuO_4$ is particularly well known as the parent compound of a hole-doped high-$T_c$ superconductor family in close proximity to a Mott insulator ground state with antiferromagnetic ordering \citep{Takagi2007}.  Whilst $\rm Sr_2RuO_4$ has the same crystal symmetry, its exotic low-$T_c$ superconducting state emerges from a Fermi-liquid metallic state \citep{Mackenzie2003}. Its 5d counterpart \SrIrO{} represents, in turn, a prototype spin-orbit Mott insulator \citep{BJKim2008,BJKIm2009,JKim2012}. 
In the presence of strong SOC, the five \ttg{} states of the $\rm Ir^{4+}$ ions with 5d$^5$ electron configuration form bands described by the effective quantum numbers \jeff=3/2 and \jeff=1/2. The latter, which is half-filled, is split already by moderate $U$ into a lower and upper Hubbard band, opening the spin-orbit Mott gap. 

The magnetic interactions in \SrIrO{} also bear a resemblance to those in $\rm La_2CuO_4$ and can be described within an antiferromagnetic Heisenberg model with an effective spin 1/2 on a quasi two-dimensional square lattice \cite{JKim2012}. The discovery of a low temperature $d$-wave gap \cite{YKim2016,Yan2015} and a splitting of the Fermi surface into so-called separated Fermi-arcs \cite{YKim2014} in electron doped \SrIrO{}, which are hallmarks of the doped cuprates \cite{Keimer2015}, have recently been reported. These findings underscore the similarity of the the low-energy effective physics of \SrIrO{} and that of the superconducting cuprates, and they encourage further research to elucidate the relationship between Mott physics and superconductivity and to search for new routes to high-\Tc{} superconductivity. 

Infrared and optical spectroscopies provide valuable information about the low-energy excitations, charge dynamics, and electron correlations in this class of materials. The in-plane conductivity spectra of single crystals of \SrIrO{} have been systematically studied \cite{Moon2008,Moon2009,Sohn2014}
and show evidence of the cooperative electron correlation and SOC effects in the presence of the orbital-dependent electron-phonon interaction.
To make further inferences about the electronic structure and underlying interactions in the layered iridate \SrIrO{}, the interplane response needs to be carefully examined; likewise in the ruthenates and cuprates\cite{Katsufuji1996,Pucher2003,Dordevic2003,Schafgans2010,Basov2005,Boris2002,Yu2008,Dubroka2011}, where valuable information about the interplane coupling, phonon anomalies, the pseudo-gap phase and precursor Cooper-pair formation has been drawn from studies of the interlayer electrodynamics.

Furthermore, $c$-axis optical conductivity data, along with the in-plane spectra, can be used to significantly constrain the model parameters for band structure calculations, such as the on-site Coulomb interaction $U$.

A thorough and reliable study of the interplane response is impeded by the small size of the currently available crystals along the $c$-axis, orthogonal to the $\rm IrO_2$ planes. The interplane optical conductivity measured on $a$-axis-oriented \SrIrO{} epitaxial films \cite{Nichols2013} is obscured by the substrate contribution and by the distorted electronic structure caused by the anisotropic biaxial strains \cite{Lupascu2014,Serrao2013}. Instead, we have used an array of high-quality and well-aligned \SrIrO{} single crystals stacked along the $c$-axis.

In this paper, we report a comprehensive ellipsometric study of the dielectric function anisotropy of \SrIrO{} over a wide range of photon energies, extending from the far-infrared (far-IR) into the ultraviolet (UV), and its interpretation based on band-structure and lattice dynamics calculations. The paper is organized as follows. Section~\ref{sec:details} describes experimental and computational details. In Section~\ref{sec:phonons}, the far-IR in- and out-of-plane phonon spectra are reported, followed by group theory analysis of the zone-center phonons and first-principles lattice dynamics calculations. The optical anisotropy of the interband transitions is discussed in Section~\ref{sec:optical}. In Section~\ref{sec:LDA} relativistic calculations, which use the local spin density approximation +U (LSDA+U) approach to account for the on-site Coulomb interaction U simultaneously with strong spin-orbit coupling, are reported in order to explain the observed anomalies and the anisotropy of the optical response. Finally, our conclusions are summarized in Section ~\ref{sec:summary}.

\section{experimental and computational details}
\label{sec:details}
High quality single crystals of \SrIrO{} were grown by a self flux method following Ref.~\onlinecite{Cao1998}. They crystallize in the {K$_2$NiF$_4$} structure with lattice parameters $a\approx\SI{5.49}{\angstrom}$ and $c\approx\SI{25.83}{\angstrom}$.
The plate-like crystals mechanically extracted from the crucible had lateral dimensions of about \SI[product-units = power]{1.5 x 2}{\milli\metre} in the $ab$-plane and thicknesses less than \SI{100}{\micro\metre} in the $c$-direction.
In order to gain a sample thickness along the $c$-axis that is sufficient for optical spectroscopy we prepared stacks of about 10 to 15 individual single crystals glued on top of each other by a minimal amount of GE-varnish.
The crystals were co-aligned according to their in-plane crystallographic axes using Laue x-ray back scattering. Subsequently one $ac$-face of the stack was polished with dry polishing paper.
The in-plane optical data were obtained from individual as-grown plate-like crystals.

We report spectroscopic ellipsometric data in a wide energy range from \SI{12}{meV} to \SI{6}{eV} over temperatures $T$ = \SI{10}{K} to \SI{300}{K}. 
For the IR range, we used home-built ellipsometers in combination with a Bruker IFS 66v/S and Vertex 80v Fourier Transform IR spectrometers. Some of the experiments were performed at the infrared beam line IR1 of the ANKA synchrotron light source at Karlsruhe Institute of Technology, Germany. Spectra in the visible and UV range were measured with a Woollam VASE variable angle spectroscopic ellipsometer equipped with an ultra-high vacuum cold-finger cryostat.

Spectroscopic ellipsometry determines the complex reflectance ratio
\begin{equation}
\rho=\frac{r_{pp}}{r_{ss}}=\tan{\Psi}e^{i\Delta}
\label{eq:}
\end{equation}
where $r_{pp}$ and $r_{ss}$ denote the reflectance of $p$ and $s$ polarized light and are given by the Fresnel equations, from which the full dielectric response $\varepsilon(\omega)$ is extracted.

For the calculations we used the low temperature experimental structural data according to Ref.~\onlinecite{Shimura1995}.
To calculate the phonon spectrum, we employed scalar relativistic density functional perturbation theory\cite{Baroni2001} as implemented in \emph{quantum espresso}\cite{Giannozzi2009}. We used ultrasoft pseudopotentials\cite{USPS}, the generalized gradient approximation\cite{Perdew1996} and set the wave function (charge density) plane wave cutoff to 80~Ry (960~Ry), respectively. The initial structure was optimized to have a stress below 0.1~kbar and residual forces per atom smaller than 0.1~mRy/Bohr prior to the lattice dynamics calculation.

The relativistic band structure calculations were performed within the local spin density approximation (LSDA) using the linear muffin-tin orbital (LMTO) method \cite{Andersen1975,book:AHY04}.
The Coulomb interaction of $\rm Ir$ 5d electrons in the presence of strong SOC was taken into account using  the rotationally invariant LSDA+U method\cite{LAZ95,YAF03}. The on-site Coulomb repulsion $U$ was varied in the range from 1.15\,eV to 2.15\,eV and Hund’s coupling $J_H = 0.65\,\text{eV}$ was fixed to the value estimated from LSDA. Thus, the parameter $\Ueff=U-J_H$, which crudely determines the splitting between the lower and upper Hubbard bands, varied between 0.5 and 1.5\,eV.
Since the calculated optical conductivity does not show any significant dependence on the actual spin orientation within the antiferromagnetic phase, we assumed collinear antiferromagnetic order in $ab$-plane with $\rm Ir$ moments aligned along the $c$-axis.

\section{phonon spectrum}
\label{sec:phonons}

\begin{figure}[b]
		\includegraphics[width=1.00\columnwidth]{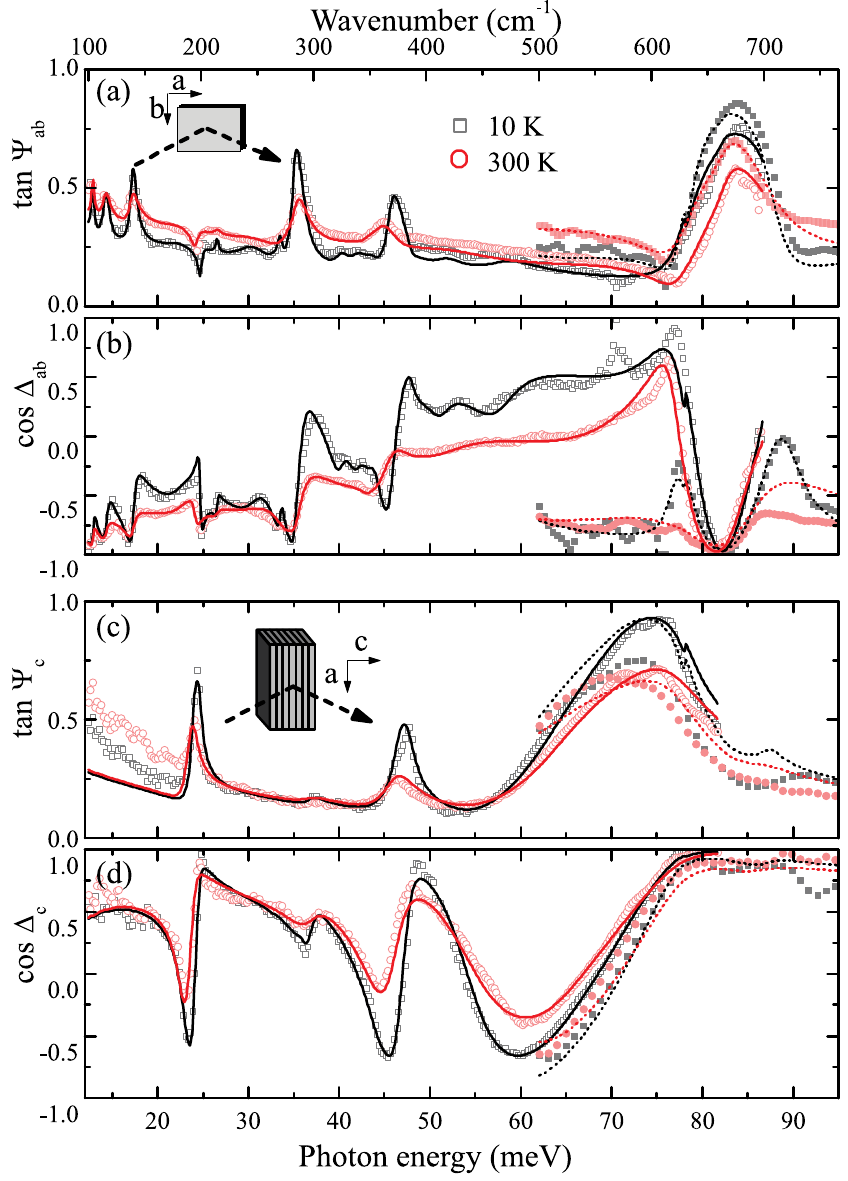}
	\caption{\label{fig:psidelta}Ellipsometric angles $\tan\Psi$ and $\cos\Delta$ in the far-infrared spectral range measured on (a,b) the $ab$-plane and (c,d) the $ac$-plane, with the $c$-axis aligned in the plane of incidence at $T=10\,\text{K}$ (grey) and $T=300\,\text{K}$ (red). Solid lines are the results of fits to model calculations involving of multiple harmonic oscillators with Lorentzian lineshapes.}
\end{figure}

\begin{figure}[t]
		\includegraphics[width=1.00\columnwidth]{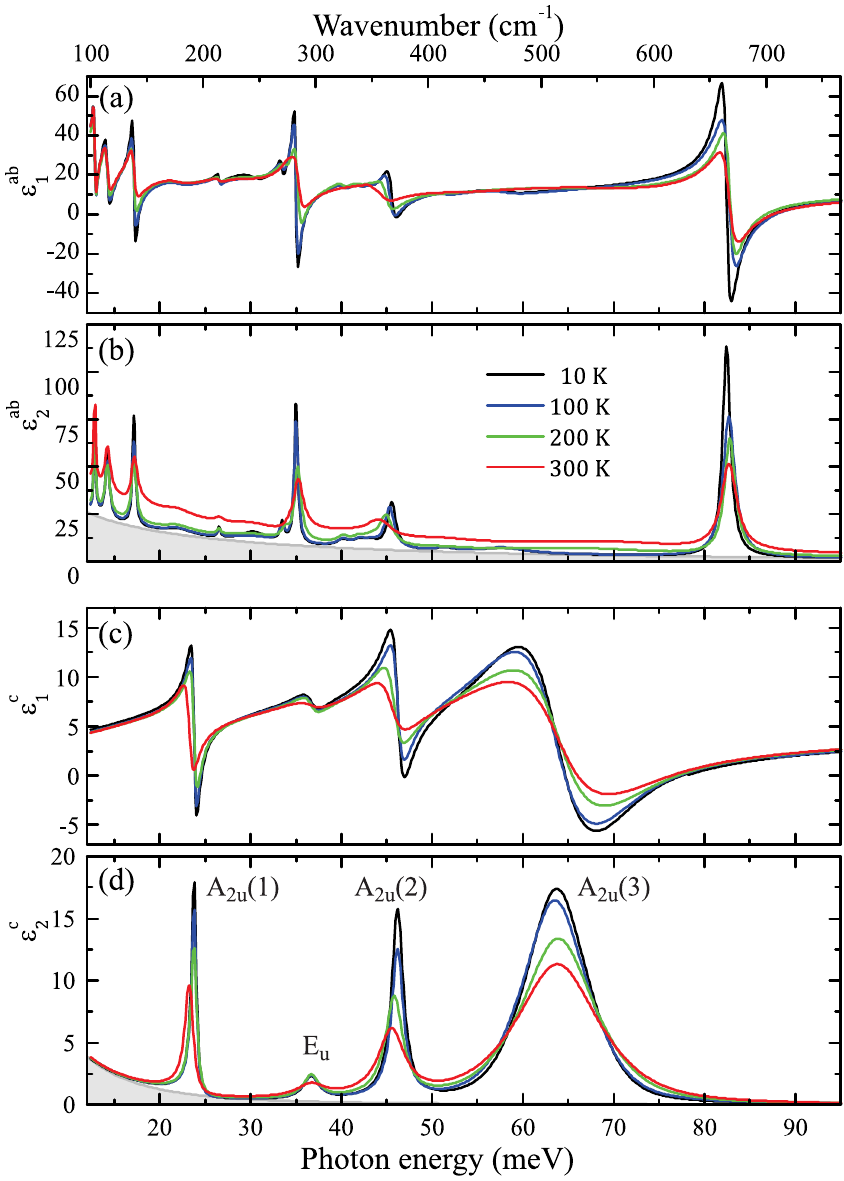}
	\caption{\label{fig:phonons}Fitted far infrared phonon spectra. Real parts of the optical conductivity $\sigma_1$ and permittivity $\varepsilon_1$ for (a,b) the $ab$-plane and (c,d) $c$-axis response, respectively, at selected temperatures.}
\end{figure}

Panels (a) and (b) of Fig.~\ref{fig:psidelta} show the ellipsometric angles $\tan\Psi(\omega,\theta,T)$ and $\cos\Delta(\omega,\theta,T)$ obtained from the $ab$-plane in the far-IR and mid-IR spectral range up to $95\,\text{meV}$ (\SI{766}{\per\centi\metre}). Panels (c) and (d) exhibit the corresponding $ac$-plane response, where the $c$-axis was co-aligned with the plane of light incidence as sketched in panel (a) and (c). Far-IR (open symbols) and mid-IR (closed symbols) data were taken at different angles of incidence and overlap in the spectral range from \SI{500}{\per\centi\metre} to \SI{690}{\per\centi\metre}.
While there is certainly a finite spread in the alignment of individual single crystals of the stacks prepared for the $c$-axis measurements there is no significant ``contamination'' from the in-plane response into the out-of-plane response as already evident from the raw data. For instance, there is no significant absorption in the $ac$-plane at the position of the highest energy $ab$-plane feature around \SI{664}{\per\centi\metre}. This underlines the validity of our stacking approach.

In the case of a system with uniaxial anisotropy the dielectric tensor $\varepsilon(\omega)$ has two complex eigenvalues --  $\varepsilon^\text{ab}(\omega)$ and $\varepsilon^\text{c}(\omega)$. In principle, two independent ellipsometric data sets on $ab$- and $ac$-faces, as presented here, allow for wavelength-by-wavelength numerical inversion of the corresponding Fresnel equations. However, due to numerical instability of the inversion process wherever the ellipsometric angles $\tan\Psi$ and $\cos\Delta$ approach their extreme values of $[0, 1]$ and $\pm 1$, respectively, which happens especially around sharp spectral features, we have fitted the full anisotropic data set at the same time by two sets of harmonic oscillators with Lorentzian lineshapes:
\begin{eqnarray}
\varepsilon^\text{ab,c}(\omega)&=&\varepsilon_1^\text{ab,c}(\omega)+ i \varepsilon_2^\text{ab,c}(\omega)\nonumber\\
	&=&\varepsilon^\text{ab,c}_\infty + \sum_{j\in \text{ab,c}}\frac{\Delta\epsilon_j\Omega_j^2}{\Omega_j^2-\omega^2-i \omega \Gamma_j}\ ,
\label{eq:Lor}
\end{eqnarray}
where $\Delta\varepsilon_j$, $\Omega_j$ and $\Gamma_j$ denote the oscillator strength, center frequency and line width of the $j$-th phonon resonance, respectively, and $\varepsilon^\text{ab,c}_\infty$ the effectively constant contribution of the high energy interband transitions to the real part of the dielectric function in the far-IR range.
For the highest energy $c$-axis phonon at \SI{515}{\per\centi\metre} we have to use a Voigt profile (that is a Lorentzian profile broadened by convolution with a Gaussian with width $\Gamma_\text{Gauss}$) to account for the anomalously large line width. This additional broadening might be caused by the stacking approach and polishing of the crystal stack.
The results of the corresponding fits are shown as solid lines in Fig.~\ref{fig:psidelta} for temperatures $T = \SI{10}{\kelvin}$ and \SI{300}{\kelvin}, respectively.
Accordingly, the fitted complex dielectric response is shown in Fig.~\ref{fig:phonons} for both the $ab$-plane and the $c$-axis also for intermediate temperatures. The best fit parameters for $T = \SI{10}{\kelvin}$ are summarized in Table~\ref{tab:fit} along with the $ab$-plane results from \citet{Moon2009}, where we find good agreement.

We unambiguously distinguish eight phonon resonances in the $ab$-plane and four in the $c$-axis dielectric response in contrast to six in-plane modes reported so far\citep{Moon2009}.
The double peak structure located at \SI{324}{cm^{-1}} and \SI{339}{cm^{-1}} develops at low temperatures only and might be related to two additional phonons. It can be clearly seen in the spectra presented by \citet{Moon2009} but is not discussed there.
The dip feature at \SI{200}{cm^{-1}} clearly seen in the raw data of the in-plane response is a result from the ellipsometric measurement scheme and correctly modeled and reproduced by the lowest energy $c$-axis phonon at \SI{192}{cm^{-1}}.
While the $ab$-plane phonons exhibit a small line width $\Gamma_j$, indicating a high single crystal quality, all four $c$-axis resonances are considerably broader. This might be attributed to the stacking procedure of many single crystals  with enhanced contribution of the near-surface regions into the $c$-axis optical response. Mechanical polishing as applied here can also induce strain effects in the surface layer which might significantly decrease the phonon lifetime \citep{Evans1974}, although the large penetration depth of far-IR radiation (of the order of several micrometers) increases significantly the bulk sensitivity, hence averaging out pure surface effects.

We find for both in- and out-of-plane response a nonzero background of absorption, which can be attributed to oxygen deficiency in the crystals \citep{Sung2016}. This might indeed have occurred under the growth conditions applied here. This background is modeled by a broad Lorentzian and shown as a shaded area for the $T=\SI{10}{K}$ case in Fig.~\ref{fig:phonons}.
It shows only moderate temperature dependence between $T=\SI{200}{\kelvin}$ and \SI{300}{\kelvin}.

\begin{figure}[htbp]
		\includegraphics[width=1.00\columnwidth]{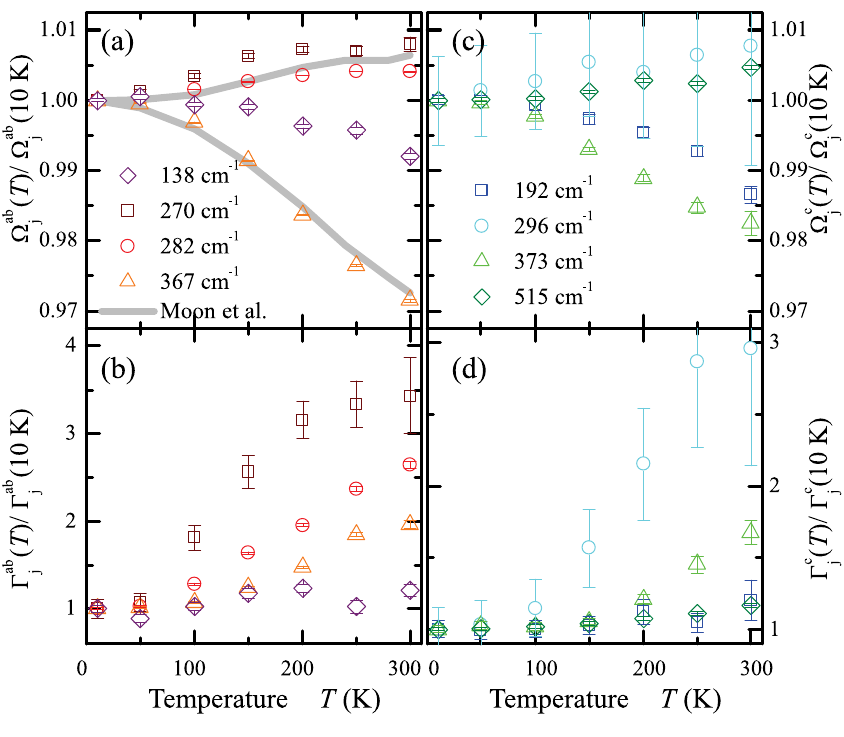}
\caption{\label{fig:parameters}Temperature dependent phonon parameters of \SrIrO{} normalized to $T={10}\,\text{K}$. (a,c) Normalized resonance frequencies $\Omega_j$ and (b,d) line width $\Gamma_j$ for $ab$-plane and $c$-axis response, respectively. The gray lines in panel (a) are reproduced from Ref. \onlinecite{Moon2009}.}
\end{figure}

\begin{table*}[t]
\caption{\label{tab:fit} Best fit results for the phonon resonances in the far-IR spectral range of \SrIrO{} at $T=10\,\text{K}$ with $\Omega_j$, $\Delta\varepsilon_j$, $\Gamma_j$  being the contribution to the static permittivity, resonance frequency and line width, respectively, according to the Lorentz oscillator model. And IR active optical zone center phonon resonance frequencies $\Omega_{j}^\mathrm{calc}$ from first-principle lattice dynamics calculations. The degree of $c$-axis polarization of the eigenvectors is given in the range $[0,1]$.}
\begin{ruledtabular}
\begin{tabular}{d{3.1} d{1.2} d{2.1} d{3.1}| d{3.3}  c  d{2.2}}
\multicolumn{4}{c}{experiment} & \multicolumn{3}{c}{calculation}\\
\\
\multicolumn{1}{c}{$\Omega_j$ (\si{cm^{-1}})} &\multicolumn{1}{c}{ $\Delta\varepsilon_j$  }& \multicolumn{1}{c}{ $\Gamma_j$ (\si{cm^{-1}})} & \multicolumn{1}{c}{$\Omega_j$  \footnote{Ref.~\onlinecite{Moon2009}} (\si{cm^{-1}}) }  & \multicolumn{1}{c}{$\Omega_{j}^\mathrm{calc}$ (\si{cm^{-1}})}&\multicolumn{1}{c}{symmetry} & \multicolumn{1}{c}{polarization} \\
\hline
		 \multicolumn{7}{c}{$ab$-plane} \\
\hline
    &      &      &     &  30&  $E_u$	&0.21\\
103 &   1  &   2.1& 102 &  81&  $E_u$	&0.08\\
115 &  1.33&   4.4& 116 &  92&  $E_u$	&0.09\\
138 &	 1.31&	 2.9& 137 & 122&  $E_u$	&0.03\\
214 &  0.07&	 2.7&     & 178&  $E_u$	&0.10\\
270 &	 0.17&	 4.4&     & 186&	$E_u$	&0.03\\	
282.5& 0.93&	 3.3& 284 & 212&	$E_u$	&0.21\\
324 &  0.13&  13  & 322 & 251&	$E_u$	&0.01\\
339 &  0.17&  16  & 338 & 298&	$E_u$	&0.15\\
367 &	 0.57&	8.8 & 366 & 406&	$E_u$	&0.00\\
664 &	 1.43&	8.6 & 663 & 645&	$E_u$	&0.00\\
    &      &      &     & 660&	$A_{2u}$&0.00\\
\hline
\multicolumn{7}{c}{$c$-axis}\\
\hline
192&	0.46&	5.2   &     & 172&	$A_{2u}$ &1.00 \\
296&	0.11& 17    &     & 323&	$E_u$	   &0.56\\
373&	0.53&	13	  &     & 374&	$A_{2u}$ &1.00\\
515&	2.6& 10 \footnote{{$\Gamma_\text{Gauss}=\SI{24}{cm^{-1}}$}}	&  & 443&	$A_{2u}$	&1.00\\
\end{tabular}
\end{ruledtabular}
\end{table*}
The temperature dependence of the resonant frequencies $\Omega_j$ and corresponding line widths $\Gamma_j$ for a representative set of in-plane and all out-of-plane modes is shown in Fig.~\ref{fig:parameters}(a-d). The resonance frequencies of the in-plane modes reproduce the reported behavior \cite{Moon2009} and the out-of-plane modes show qualitatively similar characteristics: a regular anharmonic softening by 1.5\% at \SI{300}{K} of the modes at \SI{196}{cm^{-1}} and \SI{373}{cm^{-1}}, whereas the other two exhibit a small hardening upon heating. For the $c$-axis modes as expected from the already enhanced line width at low temperatures the increase of $\Gamma_j$ with rising temperature is only moderate compared to their in-plane counterparts.

A space group analysis helps us to crosscheck our findings. 
The space group $I4_1/acd$ with the Wyckoff positions for $\rm Ir$ (8a), $\rm Sr$ (16d) and the 2 different oxygen sites (apical 16d and basal 16f) \citep{Crawford1994} allows four $A_{2u}$ and twelve doubly degenerate $E_u$ infrared active optical phonons \citep{Kroumova2003}. All modes belonging to the same irreducible representation (here either $A_{2u}$ or $E_u$) are allowed to mix in order to form the eigenmodes of the ion lattice excited by infrared photons. If an irreducible representation embodies both in- and out-of-plane polarizations this intermixing can lead to allowed phonons both in the in-plane and out-of-plane response. The large unit cell of four formula units, which adds up to 28 atoms per unit cell, and the reduced crystal symmetry makes a phonon calculation computationally expensive but indispensable for further insights on the lattice dynamics.
Therefore we compare our experimentally extracted phonon parameters with results from lattice dynamics calculations.
The set of calculated zone center phonon frequencies is summarized in Table~\ref{tab:fit}. We quantify the degree of $c$-axis polarization of each phonon eigenmode by looking at the projection $p$ of the normalized eigenvector $\vec{e}$ onto the $ab$-plane, $e_\text{ab}$, and $c$-axis, $e_\text{c}$, respectively, with $p=0.5+0.5(e_\text{c}-e_\text{ab})$ in the range [0,1].

\begin{figure}[htbp]
		\includegraphics[width=1.00\columnwidth]{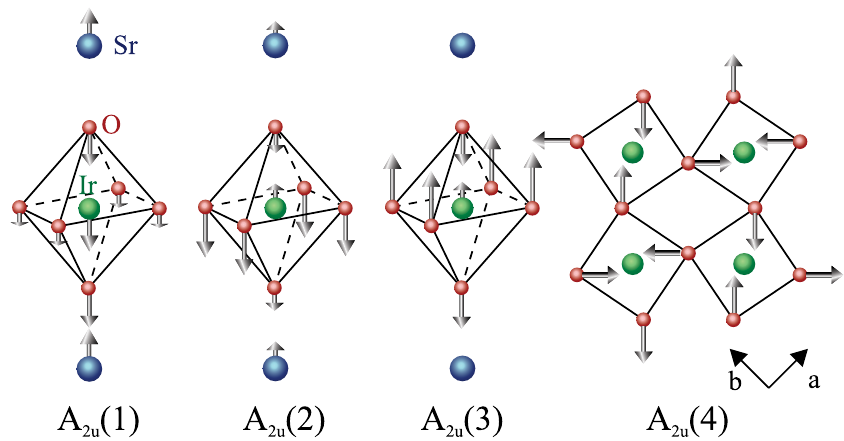}
\caption{\label{fig:pattern}Calculated eigenvectors of the infrared active optical phonon modes with $A_{2u}$ symmetry at the $\Gamma$-point of $I4/acd$ structure.}
\end{figure}

According to the calculated eigenvectors, we find that the set of four $A_{2u}$ modes actually consists of three with pure $c$-axis polarization and one polarized in the $ab$-plane. The respective mode patterns are depicted in Fig~\ref{fig:pattern}.
The $c$-axis modes $A_{2u}(1)$ to $A_{2u}(3)$ lead to octahedron bending and oscillations of the $\rm IrO_6$ octahedra against the $\rm Sr$ ions, well known from the high symmetry structure $I4/mmm$ \cite{Pintschovius1989}. 
$A_{2u}(4)$, however, involves the planar oxygen atoms only and leads to a quasi-quadrupolar mode, which stems from the backfolding of zone boundary modes (from the $M$ point of the high symmetry structure $I4/mmm$) due to the rotation of the octahedra and concomitant enlargement of the unit cell. Thus for $A_{2u}$, there is actually no intermixing of the in- and out-of-plane contributions.

In the group of the doubly degenerate $E_u$ modes, however, there is also one irreducible representation, which originates from the planar oxygen atoms, that generates a displacement in the $c$-direction. This is an IR-inactive zone boundary mode of $I4/mmm$ folded back to the $\Gamma$ point that tilts the oxygen octahedra. It mixes with other $E_u$ in-plane modes which could give rise to a non-vanishing dipole moment along the $c$-axis for several eigenmodes.
In our experimental data, however, we find in total only four out-of-plane resonances. This suggests only a weak or even absent out-of-plane dipole moment in all but one of the $E_u$ phonon modes.

With the highest phonon frequency at \SI{660}{\per\centi\metre} the total frequency range of the calculation matches quite well the experimental one. In qualitative agreement with the experimental results, we also find the highest $c$-axis phonon at considerably lower frequency  (\SI{515}{\per\centi\metre}). The considerable numerical discrepancy between the calculated and measured phonon frequencies is probably due to the fact that the material is metallic in this calculation. The phonon frequency and oscillator strength are therefore affected by charge screening, which is not present in the Mott-insulating compound.

Following this analysis and in contrast to previous work \cite{Cetin2012,Sohn2014} we assign the experimental $c$-axis modes at \SI{192}, $373$ and \SI{515}{\per\centi\metre} to be the one of $A_{2u}$ symmetry and the one at \SI{296}{\per\centi\metre} of $E_u$ type.
The in-plane modes are of $E_u$ symmetry, except of the highest energy one, that could be either related to the $A_{2u}$ quasi-quadrupolar mode or the $E_u$ mode, which we find close by in energy in the calculation.

In summary, the far-infrared phonon spectrum is consistent with group symmetry considerations and lattice dynamics calculations. Distinct in- and out-of-plane spectra prove that the sample stack can be considered as a quasi-single domain in terms of its optical response, which allows us to examine the anisotropic dielectric tensor also at higher photon energies.

\section{optical anisotropy}
\label{sec:optical}
\begin{figure*}[htbp]
		\includegraphics[width=1.00\textwidth]{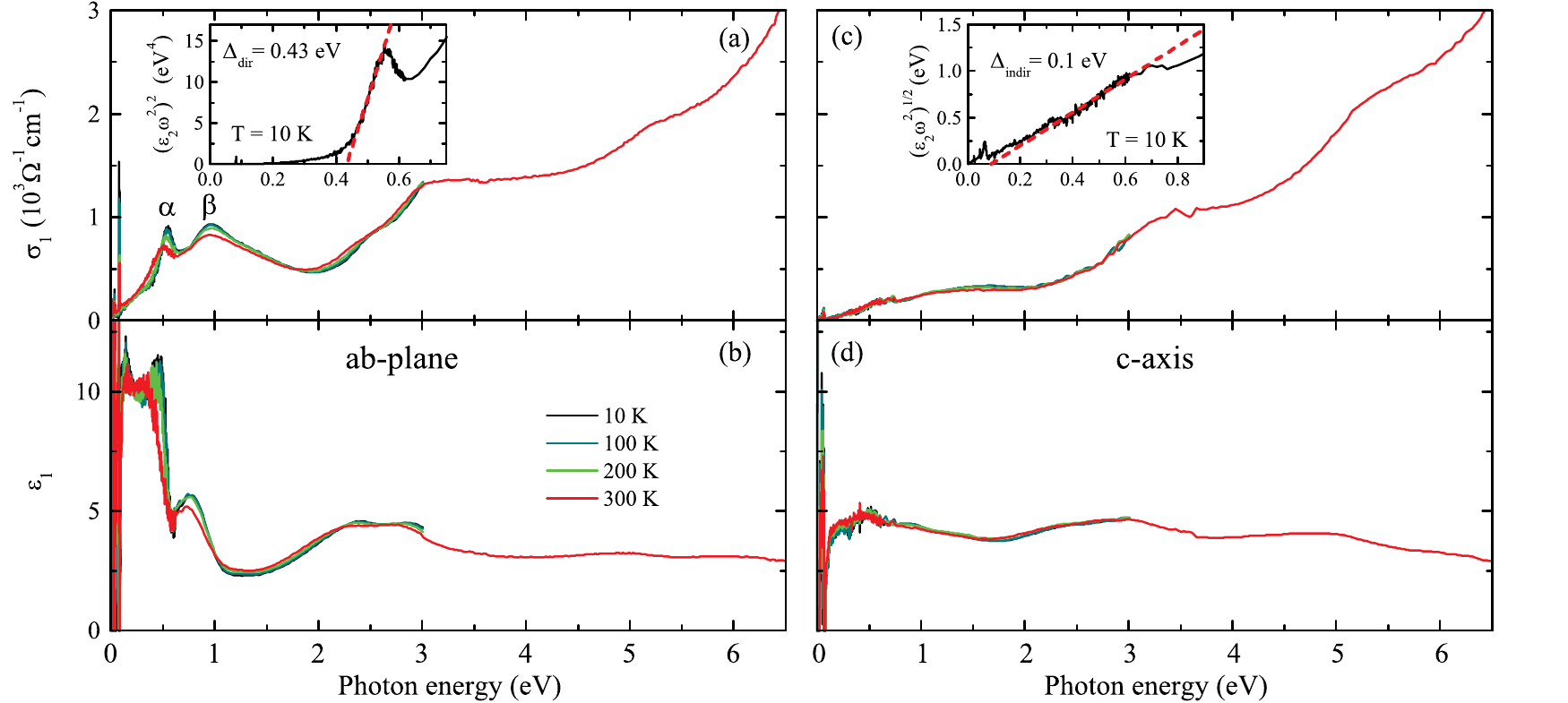}
\caption{\label{fig:fullspectra}Temperature dependence of the real parts of the optical conductivity $\sigma_1$ and dielectric permittivity $\varepsilon_1$ of \SrIrO{} in the spectral range of {0.01} to ${6.5}\,\text{eV}$. (a,b) $ab$-plane and (c,d) $c$-axis optical response. $\alpha$ and $\beta$ denote the low energy interband transitions between. Insets: (a) In-plane direct optical gap $\Delta_\mathrm{dir} = {0.43}\,\text{eV}$ extracted from $(\varepsilon_2\omega^2)^2$, (b) $c$-axis optical response $(\varepsilon_2\omega^2)^{1/2}$ and indirect optical gap $\Delta_\mathrm{indir} \approx {0.1}\,\text{eV}$.}	
\end{figure*}

Figure ~\ref{fig:fullspectra} shows the real parts of the dielectric function and optical conductivity in-plane ($\varepsilon_{a,1}$ and $\sigma_{a,1}$) as well as out-of-plane ($\varepsilon_{c,1}$ and $\sigma_{c,1}$) in the photon energy range from \SIrange{0.01}{6.5}{eV}. Since $\tan\Psi$ and $\cos\Delta$ are far from their extrema we apply the numerical inversion to correct for the anisotropy. In this energy range only moderate corrections to the absolute values of $\varepsilon$ are introduced.

First, we will focus on the in-plane response at photon energies up to \SI{6.5}{eV}. The measured spectra agree very well with the spectra previously reported in the spectral range up to \SI{3}{eV} by \citet{Moon2009} and \citet{Sohn2014}. We assign the low energy interband transitions accordingly as $\alpha$ and $\beta$. Above \SI{2}{eV} we find strong absorption setting in due to interband transitions with a plateau-like feature around \SI{3}{eV} and a further increase at higher photon energies with another shoulder around \SI{5.5}{eV}.

Following Fermi's golden rule one finds the following frequency dependence of the imaginary part of the dielectric function $\varepsilon_2(\omega)$ for photon energies just above the direct optical gap $\Delta_\mathrm{dir}$ \citep{book:Yu,Menzel2009}: $\varepsilon_2(\omega) \propto \omega^{-2}[\hbar\omega-\Delta_\mathrm{dir}]^{1/2}$. Therefore we plot $(\varepsilon_2\cdot\omega^2)^2$ in the inset of Fig.~\ref{fig:fullspectra}(a) and obtain $\Delta_\mathrm{dir} = \SI{0.43}{eV}$ at $T=\SI{10}{K}$ of the linear fit. Both the amplitude and the temperature dependence of the gap are in good agreement with the reported values \citep{Moon2009}.

The out-of-plane response, however, shows remarkably different behavior. While at high photon energies the optical conductivity is almost identical to the in-plane response with very similar characteristic shoulder features around \SI{3} and \SI{5}{eV}, the two bands $\alpha$ and $\beta$ at lower photon energies are strongly suppressed with a remaining broad, hump-like background extending down to low frequencies.
Surprisingly, the temperature dependence of the $c$-axis response is rather weak up to room temperature.
This low-energy $c$-axis optical response can be understood in terms of an indirect gap.
Phonon assisted absorption across an indirect optical gap involves essentially two processes. When the photon is absorbed an additional phonon can either be absorbed or emitted in order to fulfill energy and momentum conservation conditions. While the former strongly depends on the phonon density and is therefore strongly suppressed at low temperatures, the latter is stronger and weakly temperature dependent.
For an indirect gap of two parabolic bands one expects\cite{Menzel2009} $\varepsilon_2(\omega) \propto \omega^{-2}(\hbar\omega\pm\hbar\Omega_\mathrm{ph}-\Delta_\mathrm{indir})^2$.
As depicted in the inset of Fig.~\ref{fig:fullspectra}(b), this expression provides an excellent description of the experimental data. In this way we estimate a low temperature indirect optical gap $\Delta_\mathrm{indir}$ of about \SI{0.1}{eV}. But the absence of enhanced absorption at higher temperatures suggest additional absorption mechanisms, such as impurities, to be at play. 

A similar trend has recently been reported on thin-films of \SrIrO{} epitaxially grown along the $\langle 110 \rangle$ direction on $\rm LaSrGaO_4$ (100) substrates which also provides access to the $c$-axis response by normal incidence transmission\cite{Nichols2013}. 
Although lattice mismatch and inevitable bi-axial strain effects lead to a significant orthorhombic distortion and hence a shift in the $\alpha$ and $\beta$ bands to lower energies, the main features -- strong suppression of $\alpha$ and $\beta$ followed by an upturn and shoulder around \SI{3}{eV} -- are quite similar in both sets of materials.

In an ionic picture the origin of the optical anisotropy with respect to the $\alpha$ and $\beta$ interband transitions might be attributed to inter-site hopping of the excited electrons since on-site $d$-$d$ transitions are forbidden by the dipole selection rules \citep{Nichols2013}, but this picture neglects the spatially extended nature and hybridization of the $5d$ valence states as evident from the relatively large electronic bandwidths $W$ realized in $5d$ TMOs.
To elucidate the complex pattern of interband transitions and optical gaps we compare our experimental results with the optical conductivity from electronic band structure calculations.

\section{band structure calculations}
\label{sec:LDA}
\begin{figure}[htbp]
		\includegraphics[width=1.00\columnwidth]{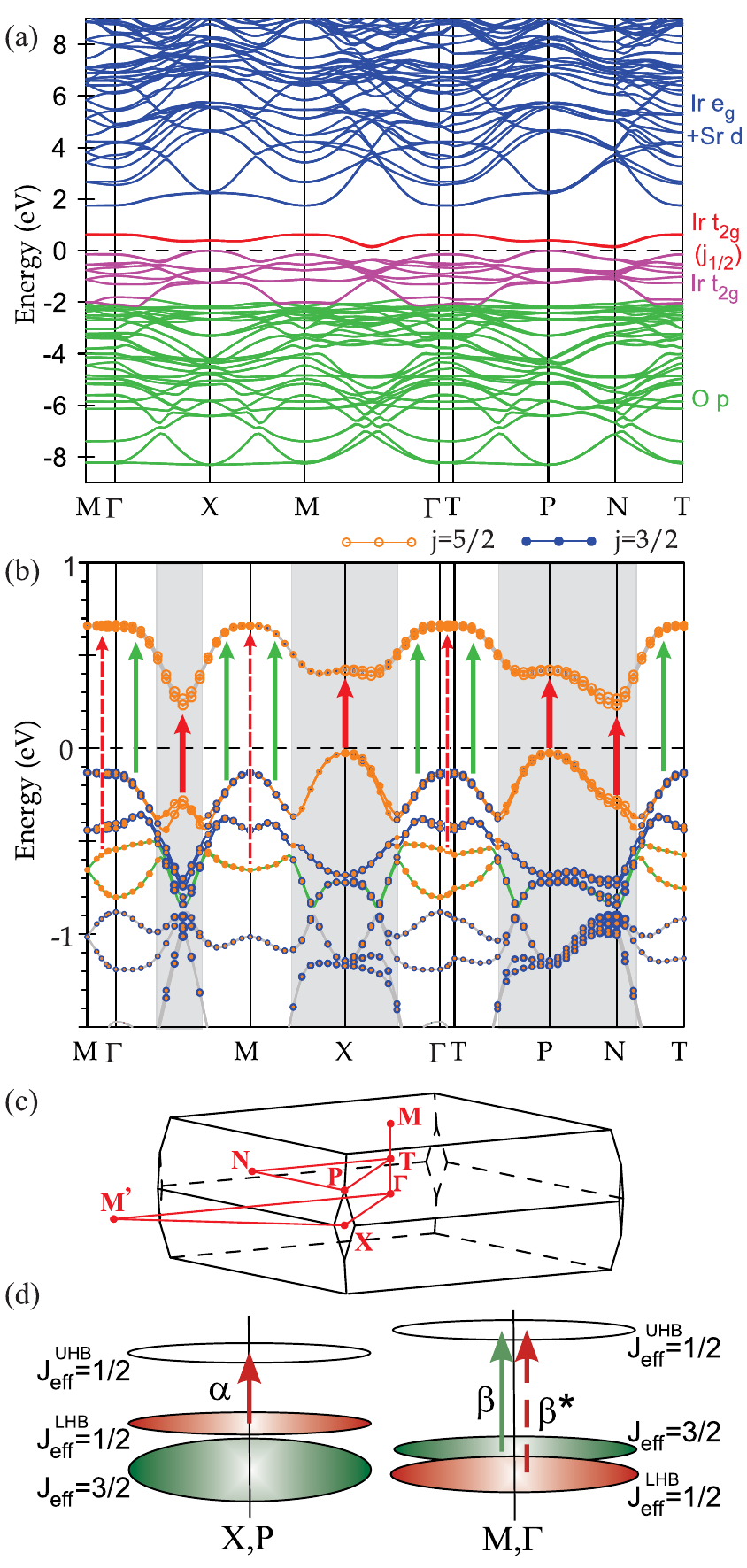}
\caption{\label{fig:bandstructure}(a) Electronic band structure and respective dominant orbital character from LSDA+U calculations with $\Ueff={1.3}\,\text{eV}$. Colors represent dominant orbital character as in the legend. (b) Enlargement of the band structure in (a). The size of the blue and orange circles is proportional to the weight of the orbital character when projected onto $d_{3/2}$ and $d_{5/2}$ states, respectively. The red and green arrows indicate $\jeff=1/2\rightarrow \jeff=1/2$ and $\jeff=3/2\rightarrow \jeff=1/2$, respectively. (c) Brillouin zone of $I4_1/acd$. (d) Sketch of the main contributions to the low energy $\alpha$,$\beta$ double-peak structure.}
\end{figure}

\begin{figure}[tbp]
		\includegraphics[width=1.00\columnwidth]{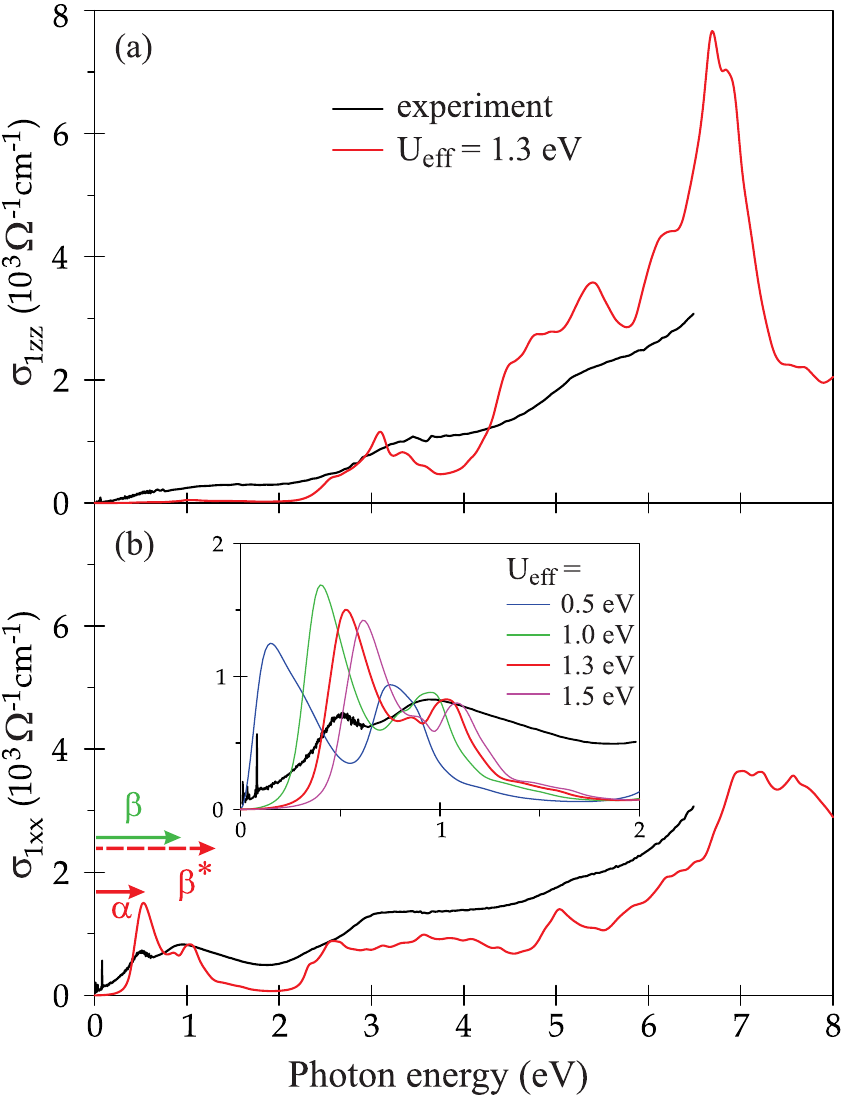}
\caption{\label{fig:LDA}Calculated (a) out-of-plane and (b) in-plane optical conductivity $\sigma_\mathrm{1,zz}$ and $\sigma_\mathrm{1,xx}$, respectively, along with the experimental spectra at $T={300}\,\text{K}$. Inset: Near-infrared in-plane double peak structure for different values of on-site repulsion $\Ueff$. $\Ueff={1.3}\,\text{eV}$ is chosen to match the position of the peaks.}
\end{figure}

The electronic band structure in the energy range of $-8$ to \SI{8}{eV} and an expansion of the region around the Fermi energy [\SI{-1.5}{eV},\SI{1}{eV}] are depicted in Fig.~\ref{fig:bandstructure}(a) and (b), respectively. In panel (b) the bands are decorated with circles proportional in size to their orbital character projected onto the basis set of the $\rm Ir$ $d_{3/2}$ (blue) and $d_{5/2}$ (orange) states.
As expected, we find the upper Hubbard band with pure $d_{5/2}$, i.e. $\jeff=1/2$, character well separated from the lower  $\jeff=1/2$ Hubbard band and the $\jeff=3/2$ states below the Fermi level [red and purple lines in Fig.~\ref{fig:bandstructure}(a)].
We adjusted \Ueff{} to match the position of the low energy $\alpha$ and $\beta$ transitions [inset of Fig.~\ref{fig:LDA}]. Hence, the direct optical band gap is naturally found to be about \SI{0.4}{eV} in our calculations. 

Before we discuss the high energy features we will focus on the analysis of the low energy $\alpha$ and $\beta$ double peak
structure of the in-plane response. The $\alpha$ band indeed stems from transitions between initial and final bands formed by pure $\jeff =1/2$ states but only from restricted sections of the $\kv$ space near zone boundaries, e.g., around the $X$-point or $P$-$N$ high symmetry line [solid red arrows and gray shaded areas in Fig.~\ref{fig:bandstructure}(b)]. 
AFM order of $\rm Ir$ moments within $ab$-plane stabilized by the on-site Coulomb repulsion $U$ causes opening of a gap near the zone boundary between two pairs of bands which are nearly degenerate in non-spin-polarized relativistic LDA calculations. These two pairs of bands show nearly parallel dispersions which insures a high joint density of states for interband transitions responsible for the $\alpha$ band.
This is in line with both previous theoretical dynamical mean-field\cite{Zhang2013} and experimental photoemission results.\cite{Wang2013} Both find the highest occupied states with $\jeff =1/2$ character around the $X$-point, too. 

The $\beta$ band located around \SI{1}{eV}, however, is more intricate.
The occupied $\jeff = 1/2$ bands exhibit rather strong dispersion with the total width of about \SI{0.8}{eV}. They cross $\jeff = 3/2$ bands so that near the Brillouin zone center along $\Gamma$--$T$--$M$ line the bottom of the $\jeff = 1/2$ bands is buried well below the top of the $\jeff = 3/2$ ones.
In agreement with previous results\cite{BJKim2008,Zhang2013} we found the $\beta$ band to have a dominant contribution from transitions with $\jeff = 3/2$ initial states [green arrows]. However, in contrast to the previous interpretation based on a
simplified band picture\cite{BJKim2008} transitions from $\jeff = 1/2$ initial bands also contribute to the optical conductivity at \SI{\sim 1.2}{eV} [dashed red arrows]. While the DMFT results\cite{Zhang2013} lead to the same conclusions, \citet{BHKim2012} favor an interpretation in terms of a Fano-type interference of the broad $\jeff = 1/2$ electron-hole continuum with an optically inactive so-called SO-exciton, i.e., a magnetically active mode found as a broad peak around \SI{0.7}{eV} in resonant inelastic x-ray experiments.\cite{JKim2012, Ishii2011} This depletes the optical excitation spectrum in that energy range which leaves the two-peak structure. Our calculations give the lower $\alpha$-band about twice as strong as the higher $\beta$-band while in the experimental spectra the strength is approximately the same for both. \citet{BHKim2012} observed a similar trend in their microscopic model calculations when considering clusters of 4 $\rm Ir$ ions. They relate this to interband mixing of $\jeff = 3/2$ and
$\jeff = 1/2$ states which reflects the itinerancy of the system, i.e., the hybridization of $\rm Ir\,d$ states via neighboring oxygen 2$p$ states.

Furthermore, our calculations allow us to analyze the full anisotropic optical response up to high energies.
Fig.~\ref{fig:LDA}(a,b) presents the real parts of the calculated optical conductivity $\sigma_\mathrm{1,zz}$ (out-of-plane) and $\sigma_\mathrm{1,xx}$ (in-plane) in the spectral range up to \SI{8}{eV} along with the experimental data taken at $T=\SI{300}{K}$.

Most noticeable is the large optical gap of about \SI{2.2}{eV} along the $c$-axis [Fig.~\ref{fig:LDA} (a)] followed by a weak band centered around \SI{3}{eV} and several stronger bands around 5.5 and \SI{7}{eV} in rather good overall agreement with the experimentally obtained $c$-axis response.
For $\alpha$ and $\beta$ we find the dipole matrix elements for $c$-axis polarization below \SI{2}{eV} either completely vanishing as for the $\jeff=1/2\rightarrow \jeff=1/2$ transitions or for $\jeff=3/2\rightarrow \jeff=1/2$ to be strongly suppressed in strength by roughly two orders of magnitude compared to the in-plane response.
Above \SI{2}{eV} interband transitions from the low lying oxygen $p$ states into the unoccupied $\jeff=1/2$ states set in. This matches the absorption edge we find in the out-of-plane response around \SI{2}{eV}.

In the in-plane response [Fig.~\ref{fig:LDA}(b)] we identify three major features. Beside the discussed double peak structure the next set of interband transitions sets in around \SI{2.2}{eV} concurrent with the $c$-axis response but at somewhat lower frequency than in the experimental spectra. From there a plateau reaches out to about \SI{5}{eV} followed by a further rise peaking at \SI{7}{eV} with approximately half the strength of the $c$-axis response.
The weak feature around \SI{5.2}{eV} seen in the experiment might find its counterpart at \SI{5}{eV} in the calculation.

Along the Brillouin zone boundaries, for example between the $P$ and $N$ points (see Fig.~\ref{fig:bandstructure}), we find an indirect optical gap of about \SI{0.2}{eV}.
This could indeed enable phonon mediated indirect transitions, which as second order processes are beyond our calculations. These transitions may contribute to the hump in the $c$-axis response as well as the strongly temperature dependent absorption edge tail and far-IR background seen in the $ab$-plane. The latter has also been shown to be relevant for $\rm Sr_3Ir_2O_7$, the narrow band gap bilayer analogue\citep{Park2014}.

\section{conclusions}
\label{sec:summary}
We demonstrate the feasibility of a stacking approach of single crystals to extract the $c$-axis optical conductivity of \SrIrO{} as proven by a distinct phonon spectrum, which is in full accordance with lattice dynamics calculations.
The observed uniaxial anisotropy in the infrared excitation spectrum is consistent with the suggested novel $\jeff=1/2$ ground state within the LSDA+U band structure calculations. The absence of the characteristic IR double peak structure in the out-of-plane response is explained by vanishing dipole matrix elements.
The additional information from the $c$-axis response severely constrains the parameter space for calculations.
Our comprehensive investigation of the optical response of a prototypical spin-orbit Mott insulator thus provides an excellent basis for experiments on doped layered iridates, which are promising candidates for exotic ground states including unconventional superconductivity.


\bibliographystyle{apsrev4-1}

%

\end{document}